PRESSURE DEPENDENCE OF THE RAMAN SPECTRUM, LATTICE
PARAMETERS AND SUPERCONDUCTING CRITICAL TEMPERATURE OF $MgB_2$

A. F. Goncharov, V. V. Struzhkin, E. Gregoryanz, H. K. Mao, and R. J. Hemley
Geophysical Laboratory and Center for High Pressure
Research, Carnegie Institution of Washington,
5251 Broad Branch Road NW, Washington D.C., 20015 USA

G. Lapertot, S. L. Bud'ko, and P. C. Canfield
Ames Laboratory and Department of Physics and Astronomy,
Iowa State University, Ames, IA 50010

I. I. Mazin
Center for Computational Materials Science, Code 6390, Naval Research Laboratory,
Washington, D.C., 20375

1. INTRODUCTION.
The newly discovered high-temperature superconductor $MgB_2$ [1] has attracted considerable interest from theoretical and experimental points of view. Theory indicates that $MgB_2$ can be treated as phonon-mediated superconductor with very strong coupling [2-5]. Transport, magnetic susceptibility, and specific heat measurements show a large isotope effect consistent with phonon mediated superconductivity [6]. Calculations show that the strongest coupling is realized for the near-zone center in-plane optical phonon ($E_{2g}$ symmetry) related to vibrations of the B atoms [2-4], which is consistent with reduced isotope effect for Mg atoms [7]. According to recent calculations [5,8], this phonon is very anharmonic because of its strong coupling to the partially occupied planar B $p_\sigma$ bands near the Fermi surface [5]. The frequency of this phonon ranges from 590 to 660 cm$^{-1}$ according to different calculations [2-5,8-10]. The phonon density of states for $MgB_2$ has been determined by neutron inelastic scattering [5,11,12], but the $E_{2g}$ mode could not be detected separately.

Raman measurements [8, 10, 3-16] indicated the presence of a broad band at 580-630 cm$^{-1}$ in general agreement with calculations for the $E_{2g}$ mode. However, the interpretation of this mode is still contradictory. Based on agreement between theory and experiment as well micro-Raman measurements on small single crystals (including polarization measurements), Refs. [10,13,15,16] assign the band to the $E_{2g}$ mode which is strongly broaden due to anharmonicity. In contrast, Refs. [8,14] suggested other interpretations, including disorder induced phonon density of state peak [14] or even a spurious feature due to an impurity phase [8]. A renormalization of the phonon frequency is expected at the transition to the superconducting phase [17]. If the background observed in the Raman scattering [8,14] has an electronic origin, one should expect to observe opening of a superconducting gap below $T_c$. The reports on this are so far inconsistent. On the other hand, a multigap structure of the order parameter, as suggested in Ref. [17], can make the gap opening much less detectable.

Pressure is an important variable that can be used to tune physical properties and compare the results with theoretical predictions. Pressure effects on superconductivity studied up to 1.84 GPa [18] and 0.5 GPa [19] under hydrostatic conditions show a decrease of $T_c$ at a rate of 1.6 K/GPa and 1.11 K/GPa, respectively. In contrast, the study of Monteverde et al. [20] show a variety of different shapes and slopes depending on loading conditions etc. This can be understood by anisotropic stress dependence of $T_c$. Very recently, a new set of data on $T_c$ versus P in nearly hydrostatic conditions has been produced in our group [21]. The pressure dependence of $T_c$ has been proposed to be mainly due to an increase in phonon frequency [22]. This conclusion was based on an assumption of largely pressure independent electron-phonon matrix element *I*, empirically predicted shift of the Raman phonon, theoretical calculation of electronic density of states and equation of state [22,23]. With new experimental data on equation of state and pressure dependence of Raman frequency, these theoretical predictions can be tested on a more solid ground. Compressibility data have been obtained by neutron diffraction (to 0.62 GPa) [24] and synchrotron x-ray diffraction (to 6.15 GPa [25] and 8 GPa [23]). We determined the equation of state to 12 GPa in purely hydrostatic conditions and also the pressure dependence of the Raman spectra to 14 GPa [13]. In this paper we present an overview of our Raman and x-ray data and compare them with other observations. New results for the pressure dependence of $T_c$ to 14 GPa are also presented along with the analysis, which allow us to obtain the electron-phonon coupling to compression $\Delta V/V_0$ of 8%.

2. EXPERIMENTAL.
Samples of Mg $^{10}B_2$ (with isotopically pure $^{10}B$) were similar to those used in Refs. [6,26]. They are essentially in powdered form consisting of aggregates of 30-50 μm linear dimensions, which is ideal for high-pressure experiments. Our experiments were done with various types of diamond anvil cells. In Raman experiments a long piston-cylinder cell was used and Ne served as a pressure transmitting medium [27]. Synthetic ultrapure diamonds were used as anvils to reduce background fluorescence. Raman scattering was excited in a $145^0$ geometry (see Ref. [28]) to reduce further the background from diamond Raman and that originating from spuriously reflected light. The spectra were recorded with a single-stage spectrograph equipped with a CCD detector and holographic notch filters (150-5000 cm$^{-1}$), although occasional measurements were also done with a conventional triple spectrometer to cover the lower frequency range. Spectra were excited by different lines of Ar-ion laser (457-514.5 nm). Most of Raman measurements were performed at room temperature, except for one low-temperature (to 12 K) excursion at approximately 8 GPa. In this case, a bath He cryostat (Cryo Industries of America) with special long-working distance optics has been used.

X-ray diffraction was measured with a symmetric diamond cell in an energy-dispersive configuration at beamline X17C of the National Synchrotron Light Source with $2\theta=10^0$ [29]. In the x-ray experiment we used helium as a pressure transmitting medium, which is purely hydrostatic to 12 GPa. Pressure was determined by the standard ruby fluorescence technique. All measurements were performed at room temperature.

High-pressure measurements of $T_c$ were done with Be-Cu nonmagnetic diamond anvil cell [38] with a flat tip 400 μm diameter anvils. A flake of $Mg^{10}B_2$ approximately 40 μm in diameter and 10 μm thick was loaded into a hole approximately 200 μm in diameter, prepared in a nonmagnetic NiCr(Al) gasket with He as the pressure transmitting medium. Pressure was increased at 40 - 50 K, and determined by the standard ruby fluorescence technique. Our results show very good agreement with hydrostatic data, so we believe that nonhydrostatic effects in our experiment are negligible. We warmed the sample to room temperature during this experiment at 10 GPa and kept it at room temperature for few days. This did not have any effect on the observed pressure dependence of $T_c$, so we conclude that there is no substantial effect on $T_c$ due to thermal cycling. We also performed a nonhydrostatic experiment up to 25 GPa without pressure medium. However, the observed pressure dependence was different from the run in He pressure medium, and resembled the data for powdered sample obtained by Monteverde et al. [20].

3. RESULTS AND DISCUSSION.
Figure 1 presents the Raman spectra at different pressures. The broad band observed is a Raman excitation as shown by changing the laser wavelength and by anti-Stokes measurements. It was also checked that the signal originates from $MgB_2$ because identical spectra were recorded by separate micro-Raman measurements from individual micron-size grains (shown in Fig. 1 as the 0 GPa spectrum). Also, the Raman spectra contain a wide unstructured background component. This may be analogous to the similar effects observed in cuprate HTSC materials [31], where the background has an electronic origin; on the other hand, electronicaly $MgB_2$ is a simple metal, where noticeable electronic Raman scattering is not expected below the plasmon frequencies, as opposed to the "bad metal" cuprates. Pressure leads to an increase in the frequency of the broad band without any appreciable change of its shape. The spectra can be fitted reasonably well with a combination of a linear background and a Gaussian peak (fitting with a Lorenzian peak gives almost the same results and similar quality of fitting). The frequency determined by this procedure is plotted as a function of pressure in Fig. 2. We increased and decreased pressure several times and found that the effects observed are fully reversible. The pressure dependence is linear within the accuracy of the experiment. Pressure dependence of the mode damping is much smaller, 10%/15 GPa versus 25% for the shift (see inset in Fig. 2), so that the *relative* linewidth drops by ~17% over 15 GPa, suggesting some reduction in the electron-phonon coupling constant with pressure.

We also performed a separate experiment with no transmitting medium in order to examine whether nonhydrostatic conditions have any effects on the observed Raman band. We find that in this case the second Raman peak at about 800 cm$^{-1}$ gradually evolves with pressure and takes over in intensity above 10 GPa. This observation corroborates very well with that reported in Ref. [8] and should be ascribed to effects related to compacting of powder samples under nonhydrostatic conditions. In addition, some of our x-ray diffraction powder patterns showed a splitting of (002) reflection under similar conditions.

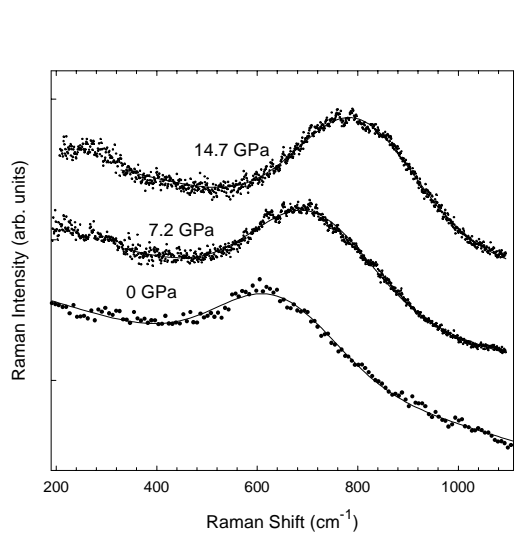 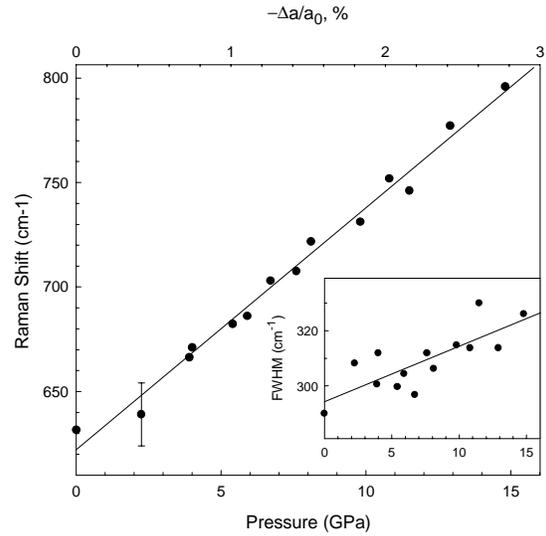

Fig. 1. Raman spectra of MgB$_2$ at selected pressures. Spectra are shifted vertically for clarity. Points are experimental data and solid lines represent the phenomenological fits (see text) to the spectra in the appropriate spectral range. The excitation wavelength was 514.5 nm.

Fig. 2. Raman frequency as a function of pressure and relative compression of the *a* lattice parameter (upper scale). Points are experimental frequencies determined from the phenomenological fits of the spectra. The solid line is a linear fit. The inset shows the pressure dependence of the damping obtained by the same fitting procedure.

We made a low-temperature excursion attempting to study temperature effects and a variation of spectra at the transition to the superconducting state. During this run, the pressure varied from 4 to 8.5 GPa (mainly at 100-300 K) because of thermal contraction of our diamond cell. Raman spectra at different pressures are presented in Fig. 3. No dramatic changes were found at low temperature including the range below the transition to the superconducting state (<30 K). The only noticeable effect is a relatively small reduction of the linewidth in the superconducting state (Fig. 4), while no measurable change of frequency (see inset) was detected (c.f., Ref. [17]).

Factor-group analysis predicts for $MgB_2$ (space group P6/mmm, Z=1)
$\Gamma = B_{1g} + E_{2g} + A_{2u} + E_{2u}$ zone center optical modes, of which only $E_{2g}$ is Raman active. Thus, it is natural to assign the band observed at 620 cm$^{-1}$ at ambient conditions to the $E_{2g}$ mode (see also Refs. [10,15,16]). The frequency agrees well with theoretical calculations [5,8,10]. The anomalously large linewidth (FWHM=300 cm$^{-1}$) can be ascribed to large electron-phonon coupling [10], which will be described below.

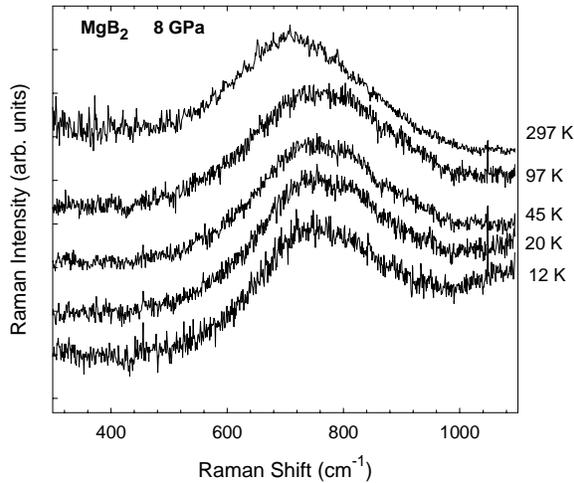

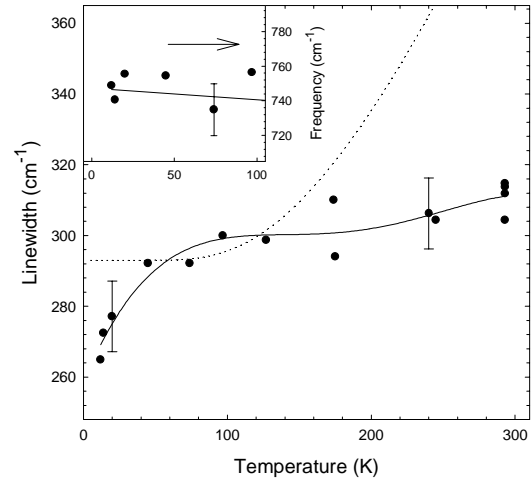

Fig. 3. Raman spectra of $MgB_2$ at a nominal pressure of 8 GPa and different temperatures. A background has been subtracted from the spectrum at 297 K.

Fig. 4. Temperature dependence of the linewidth of the $E_{2g}$ mode measured at pressure of 7-8 GPa. The inset shows the temperature dependence of frequency at 8 GPa. The dotted line is a theoretical curve calculated assuming that broadening is due to decay in two acoustical phonons (e.g. J. Menendez & M. Cardona, Phys. Rev. B29 (1984) 2051.)

Attempts to ascribe the mode observed to impurities [8] contradict polarization measurements on small single crystals [13,15,16]. These observations also rule out another alternative interpretation of the 600 cm$^{-1}$ peak, disorder-induced scattering, as proposed in Ref. [14].

The experimental pressure dependences of the lattice parameters determined by x-ray diffraction are shown in Fig. 5. Our data are in good agreement with Refs. [23,24], while the results of Ref. [25] show systematically larger lattice parameters and yet comparable compressibility. We calculated the bulk modulus $K_0$ assuming "normal" behavior and $K_0'=4$, which is typical for covalent and metallic bonding [32] (our data do not allow us to fit data with two parameters $K_0$ and $K_0'$). The result is 155(10) GPa, in good agreement with Refs. [22-24]. The axial compressibilities are $\beta_a$=0.0016(2) GPa$^{-1}$ and $\beta_c$ =0.0030(2) GPa$^{-1}$.

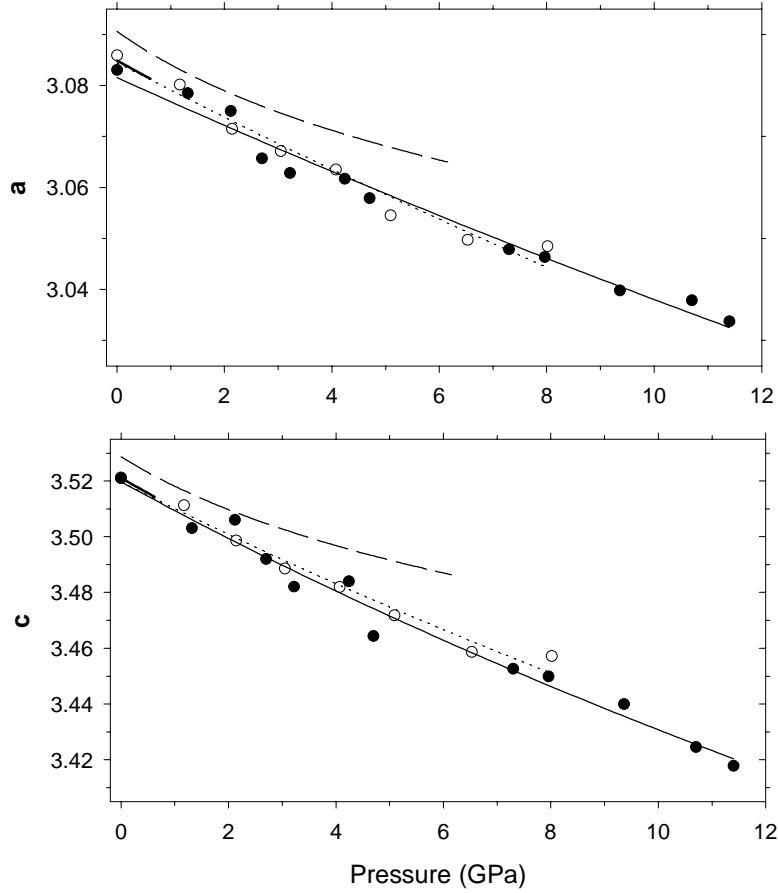

Fig. 5. Experimental pressure dependences of the lattice parameters. Filled circles with solid line (Murnagan fit) are our data; thick solid lines are from Ref. [24]; dashed lines are from Ref. [25]; open circles and dotted lines are from Ref. [23].

The mode Grüneisen parameter $\gamma_0$= $K_0$dln$\nu$/dP (calculated at 0 GPa) determined from our data is 2.9±0.3. In the case of anisotropic crystals it is more appropriate to scale the frequency shift of in-plane mode with the variation of interatomic bond distance or lattice parameter $a$ [33]. The corresponding component of the Grüneisen parameter ($\gamma_{0a}$=dln$\nu$/3dln$a$) is 3.9±0.4. These values are substantially larger than those expected for the phonon in a compound with covalent bonding [34], which should be dominant for this mode, where typically, $\gamma_0$ does not exceed 2. For example, for graphite $\gamma_0$=1.06 [33] and for iron (i.e., with metallic bonding) $\gamma$=1.7 [28]. Larger $\gamma$'s are normally related to increased anharmonicity of the mode [35]. This can also be a consequence of a soft mode behavior when the system is approaching (or departing from) a structural instability (e.g.

Ref. [36]). Large γ's were also observed in a vicinity of electronic topological phase transition [37].

We believe that the data available strongly suggest the first-order phonon interpretation because of the agreement with the calculated frequency [5,10] and linewidth. Theoretical calculations suggest a scenario with the $E_{2g}$ phonon strongly coupled to electronic excitations [5]. Our data show a very broad, strongly pressure dependent excitation, which also agrees with this interpretation because the unusually large Grüneisen parameter $\gamma_0$ may indicate the proximity of $MgB_2$ to electronic topological phase transition (c.f. Ref. [37]). The inferred large [5,17] anharmonicity is likely to be induced by nonlinear coupling with electronic degrees of freedom[17], which explains its relatively small variation with pressure and temperature (figs. 2 and 4), in comparison to "conventional" anharmonicity (see e.g. Ref. [38]). Within this picture, our data favor the coupling of the $E_{2g}$ phonon to the electronic subsystem.

In Fig.6 we show $T_c$ as a function of pressure, as well as temperature scans at selected pressures. The onset of $T_c$ can be reliably identified with accuracy 0.2-0.4 K (depending on the actual quality of the experimental data). The pressure-induced change in $T_c$ is -1.1 K/GPa near ambient pressure, which is close to the pressure derivatives reported from hydrostatic measurements [18,19,39]. We estimate that $\partial(T_c/T_{c0})/\partial(V_0/V)=4.2$ using bulk modulus 155 GPa. This can be compared with 4.16 obtained in [19,39] for $^{11}B$ isotope.

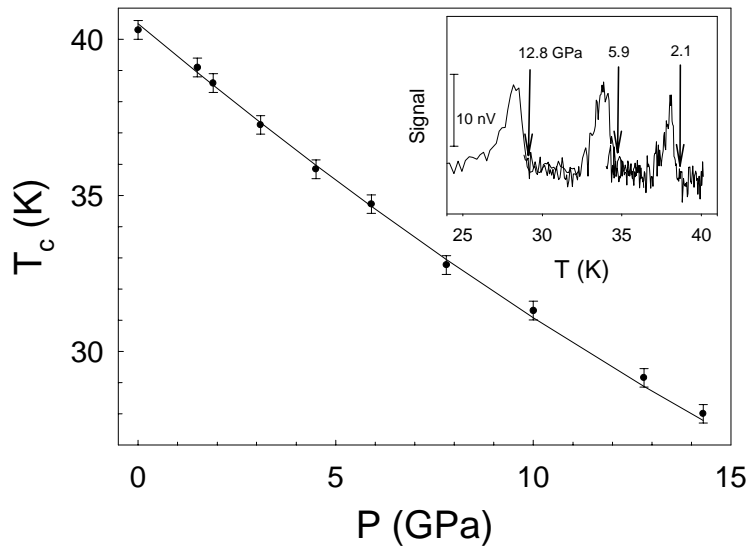

Fig. 6. Pressure dependence of $T_c$ in a He medium. $T_c$ is identified as the onset of the magnetic signal [30]. The onset is marked by arrows, as illustrated in the inset for selected pressures. The solid line is a guide to the eye.

One can analyze $T_c$ in terms of the McMillan formula [40], as modified by Allen and Dynes [41]

$$T_c = \frac{\omega_{\ln}}{1.2}\exp\left(-1.04\frac{1+\lambda}{\lambda-\mu^*-0.62\lambda\mu^*}\right), \quad (1)$$

where $\omega_{\ln}$ is the logarithmically averaged phonon frequency, $\lambda$ is the electron-phonon coupling constant, and $\mu^*$ is the Coulomb pseudopotential. The electron-phonon coupling constant can be written as

$$\lambda = N(0)<I^2>(M<\omega^2>)^{-1} \quad (2)$$

where $N(0)$ is the density of electron states at the Fermi level, $<I^2>$ is the square of electron-phonon matrix averaged over the Fermi surface, M is the mass of the atom, and $\omega^2$ is the average square of a characteristic phonon frequency [40].

We use Eq.1 to calculate $\lambda$, assuming $\omega_{\ln} \propto (V_0/V)^\gamma$ and taking the experimental $\gamma=2.9\pm0.3$, as determined from Raman measurements (assuming the variation in Grüneisen parameter in this pressure range is negligible). The resulting values of $\lambda$ are shown in Fig. 7. We use the ambient pressure $\omega_{\ln}$ (corrected for isotope shift), as well as Coulomb pseudopotential $\mu^*=0.13$ from Ref. [17]. In our analysis we assume that $\mu^*$ does not depend on pressure (in principle, lattice hardening should lead to some increase of $\mu^*$, but this is a relatively weak effect). We do not find any anomalies in $T_c(P)$ dependence to 14 GPa. We estimated the pressure dependence of the Hopfield parameter

$$\eta = N(0)\langle I^2 \rangle \propto (V_0/V)^\varphi \qquad (3)$$

using $\langle \omega^2 \rangle \propto (V_0/V)^{2\gamma}$. Here $\varphi$ is an empirical parameter, which accounts for pressure dependences of both N and I. Our data give $\varphi = 2.3 \pm 0.6$. The variation of $N(0)$ is generally given as $1/t$, where t is the p-p hopping parameter. Using Harrison's canonical scaling, one can estimate that $t \sim 1/d^3 = V_0/V$, (for isotropic compression, d is B-B bond length). However, the $p_\sigma$ and the $p_\pi$ bands move with pressure with respect to each other [42], thus inducing charge transfer between these two bands. We performed LMTO calculations, similar to those described in Ref. 2 and using the experimental lattice parameters, to compute the partial density of states of the two $p_\sigma$ bands. We found that it is proportional to $d^{2.8}$, rather close to the canonical-scaling estimate. Due to anisotropy in compressibility (*c* axis is more compressible than *a* axis) we estimate $V \sim d^{3.8}$, instead of $V \sim d^3$. Thus, $N(0)$ contributes (-0.74) to $\varphi$, and $\langle I^2 \rangle \propto (V_0/V)^{3.0(6)}$. There are two alternative approaches to calculations of $\langle I^2 \rangle$. In the Rigid Muffin Tin Approximation (e.g., Ref. 2), $\langle I^2 \rangle$ is a quasi-atomic property and does not depend on $V_0/V$; within the tight-binding, $\langle I^2 \rangle$ is related to the derivatives of the band energies with respect to the ionic displacement; thus $I \propto \partial t/\partial d \propto d^{-4} \propto (V_0/V)^{1.05}$ (again, we use $V \sim d^{3.8}$), $\langle I^2 \rangle \propto (V_0/V)^{2.1}$, in fair agreement (given all the simplifications used in the way) with $(V_0/V)^{3.0(6)}$ deduced above. We note that the covalent character of bonding in the B planes [3,43] does suggest that the tight-binding approach is more appropriate than the Rigid Muffin Tin one.

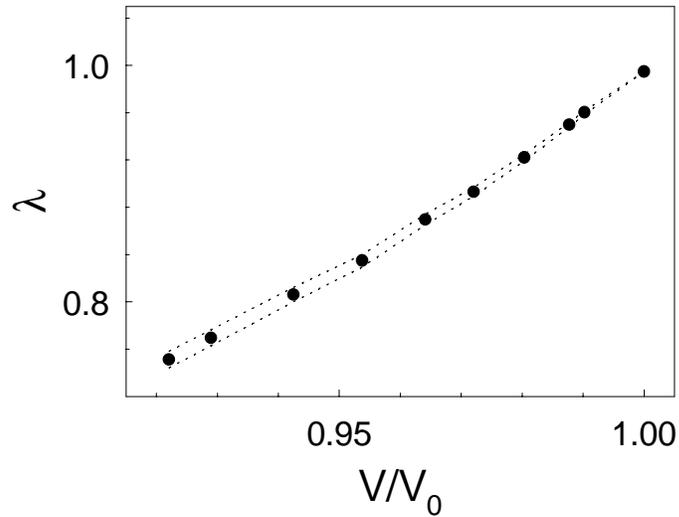

Fig.7. Electron-phonon coupling derived from $T_c(P)$ data. Solid circles are for $\gamma=2.9$. Dashed lines were calculated with $\gamma=2.6$ (upper) and $\gamma=3.2$ (lower).

Deemyad et al. suggested recently [39] that the pressure dependence of $T_c$ comes predominantly from the lattice stiffening. We also find that this effect is more important than the pressure dependence of the Hopfield parameter $\eta$, but we do not find such a strong disparity between the two factors, as Deemyad et al. In fact, they assumed that $\varphi = 1$ and ascribed all remaining pressure dependence to the lattice stiffening, and found that they need a Grüneisen parameter of 2.36 to produce the required pressure dependence of $T_c$. However, the actual Grüneisen parameter, as measured in our experiments, is 2.9. Correspondingly, a considerably stronger pressure dependence of the Hopfield parameter is required to offset the stronger lattice stiffening, as quantified by our $\varphi = 2.3$. However, there are also some discrepancies between our results and those from Ref. 39 in the pressure dependence of $T_c$ itself. We compare our data with Ref. [39] in Fig. 8. We find a linear dependence of $T_c$ on volume, as compared to the sublinear dependence from Ref. [39]. The difference in bulk modulus values is insufficient to explain the observed trend, which gets even larger if we use bulk modulus 150 GPa (closer to 147.5 GPa used in [39]) instead of 155 GPa. Part of the discrepancy can be explained by the procedure used to determine $T_c$: it was taken as the midpoint of the magnetic transition in [39], whereas we have used the onset of the magnetic transition. However, we do not expect that the difference may be more than 0.5 K, because the width of the transition was reported to be less than 0.9 K in [39]. This brings the observed differences close to the experimental accuracy of the measurements. Sample-dependent

$T_c(P)$ is also a possibility. Similar measurements on isotopically pure samples by both groups may

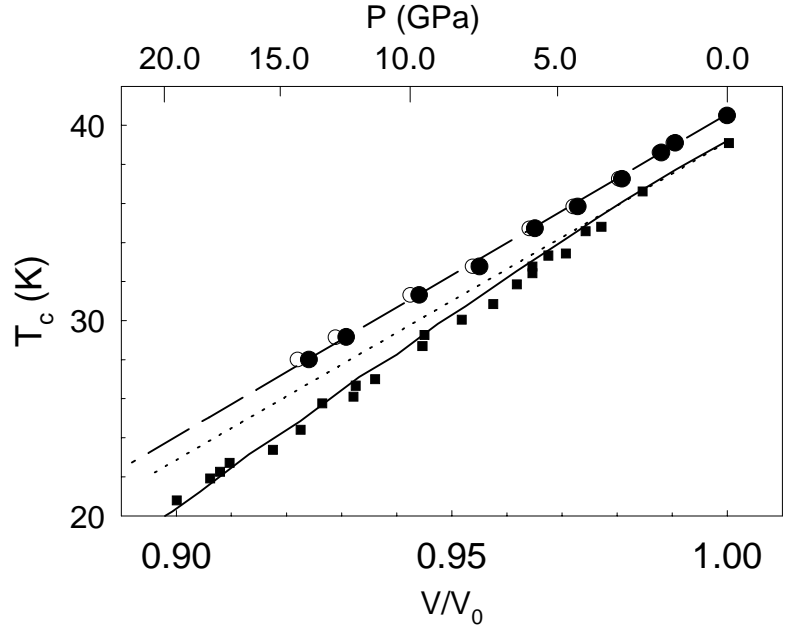

Fig. 8. Our $T_c$ data versus volume: solid circles for a bulk modulus of 155 GPa; open circles for bulk modulus 150 GPa; dashed line is a linear fit. The squares are from Ref . [39], dotted line is the linear fit to low pressure data from Ref. [39]; solid line is $T_c=T_{co}(V/V_0)^{4.16}$.

exclude possible systematic errors, which will help to look for a pressure-dependent isotope effect.

In conclusion, we have observed a strongly broadened Raman band of $MgB_2$ that shows anomalously large pressure dependence of its frequency. This band and its pressure dependence can be interpreted as the $E_{2g}$ zone center phonon, which is strongly anharmonic because of coupling to electronic excitations. The pressure dependence of $T_c$ was measured to 14 GPa and can be explained only when a substantial pressure dependence of the Hopfield parameter $\eta=N(0)<I^2>\propto (V_0/V)^{2.3(6)}$ is taken into account.

The authors are grateful to P. Dera, J. Hu for help with the experiments. The authors also thank O. Gulseren and T. Yildirim for useful discussions. Work at Ames Laboratory was supported by the Director of Energy Research, Office of Basic Energy Sciences, U.S. Department of Energy. Work at NRL was supported by the Office of Naval Research. We acknowledge financial support of CIW, NSLS, NSF and the Keck Foundation.